\documentstyle[aps,epsf,twocolumn]{revtex}
\input epsf.sty
\begin{document}
\draft
\flushbottom
\twocolumn[
\hsize\textwidth\columnwidth\hsize\csname @twocolumnfalse\endcsname

\title{Superconducting phase coherence in striped cuprates}
\author{A. H. Castro Neto}

\address{Department of Physics,
University of California,
Riverside, CA, 92521}
\date{\today}
\maketitle
\tightenlines
\widetext
\advance\leftskip by 57pt
\advance\rightskip by 57pt

\begin{abstract}
We study the problem of phase coherence
in doped striped cuprates.
We assume the stripes to form a network of one-dimensional Luttinger
liquids which are dominated by superconducting fluctuations and pinned
by impurities. We study the dynamics
of the superconducting phase
using a model of resistively shunted junctions which
leads to a Kosterlitz-Thouless transition. We show
that our results are consistent with
recent experiments in Zn-doped cuprates. We also explain
the scaling of the superconducting critical temperature $T_c$
with the incommensurability as seen in recent neutron scattering
experiments and predict the behavior of $H_{c2}$.
\end{abstract}
\pacs{PACS numbers:74.20.Mn, 74.50.+r, 74.72.Dn, 74.80.Bj}

]
\narrowtext
\tightenlines

It is already well established that cuprates have a strong tendency
towards phase separation \cite{review}.
Macroscopic phase separation has been observed in
La$_{2}$CuO$_{4+\delta}$ \cite{hammel}. In other materials,
however, phase separation is frustrated, and one
observes the formation of domain walls or stripes. Stripes have been seen
experimentally in La$_{2-x}$Sr$_x$NiO$_{4+y}$ \cite{nickel}. Magnetic
susceptibility measurements, nuclear quadrupole resonance
and muon spin resonance \cite{msr}
indicate formation of domains in La$_{2-x}$Sr$_x$CuO$_{4}$.
This picture is not inconsistent with neutron scattering experiments
in YBa$_2$Cu$_3$O$_{7-\delta}$ \cite{tran}.
More recently a direct evidence for stripe formation was given
in neutron scattering in
La$_{1.6-x}$Nd$_{0.4}$Sr$_x$CuO$_{4}$ \cite{tranquada}.
Phase separation
can be frustrated by the long range Coulomb repulsion between the holes
\cite{steve,rapid} or disorder induced by dopants \cite{nihat}.

Superconducting
cuprates are naturally disordered because the charge carriers
have their origin on doping. In this case holes and impurities
have opposite charge which leads
to hole-impurity attraction. Localization
of holes close to O-impurities has been seen in La$_{2}$CuO$_{4+\delta}$
\cite{hammel}. {\it Ab initio} calculations seem to imply that a large
percentage of the holes can be localized in these materials \cite{martin}.
Another way to localize charges in the CuO$_2$
planes has antiferromagnetic origin. Zn, for instance,
when substituted on the Cu sites, hybridizes poorly with O atoms.
Thus Zn breaks local antiferromagnetic bonds.
In this case the holes can take advantage of the
smaller number of bonds and localize close to the Zn sites.
Therefore, while phase separation can lead to stripe formation,
impurities can
cause stripe pinning. Furthermore, one expects
strong magnetic distortions around
the impurities \cite{walstedt,ng}.

In our picture the stripes are quasi-one-dimensional regions
of the CuO$_2$ planes where the holes are segregated.
We have recently proposed a model \cite{euro} that explains 
the dependence of the 
antiferromagnetic-paramagnetic phase transition on doping \cite{prl} and
recent neutron scattering experiments by Yamada {\it et al.} \cite{yamada}
in La$_{2-x}$Sr$_x$CuO$_{4}$. In these experiments a
commensurate-incommensurate transition is observed as a function of doping.
It is well known that the incommensurate magnetic peaks are seen at
$(\pi/a \pm \epsilon,\pi/a)$ and $(\pi/a,\pi/a \mp \epsilon)$ where $\epsilon$
depends on doping. In the stripe picture one has 
$
\epsilon = \pi/\ell
$
where $\ell$ is the inter-stripe distance \cite{euro}.

\begin{figure}
\epsfysize5 cm
\hspace{0cm}
\epsfbox{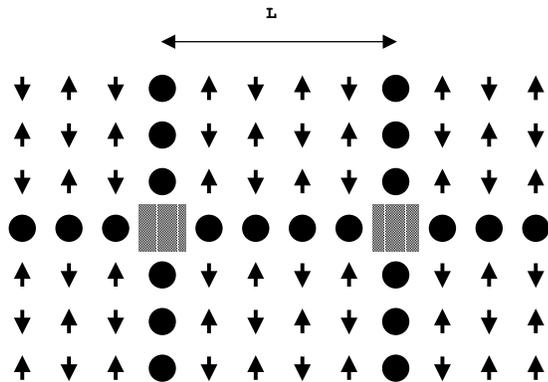}
\caption{Geometry of the problem. The antiferromagnet
(shown with $\uparrow,\downarrow$) in an anti-phase domain
configuration together with stripes (as black circles) and
impurities (gray squares). Observe that the distance between
impurities $L$ is the inter-stripe distance $\ell$.}
\label{imp}
\end{figure}

Since stripes are one-dimensional objects
they cannot show true long range superconducting order \cite{boso}. This is
only possible if the stripes interact with each other
by exchanging Cooper pairs. There are a few ways the stripes can interact.
One of them is by a direct exchange of Cooper pairs
via stripe fluctuations. This is a dynamical
Josephson effect \cite{bala}. Another possibility is
due to stripe crossing in the
presence of impurities.
In this paper we propose a scenario where the holes are localized
close to the impurities and these ``lakes" of holes are connected
amond themselves by ``rivers" or stripes forming a network.
See Fig.(\ref{imp}). We show that stripes
carry a Josephson and a normal current. While the normal current
dissipates, the Josephson current can lead to the exchange
of Cooper pairs via the lakes of holes. Moreover, since there
is an accumulation of holes close to the impurities, one expects
charging effects to play a role in the problem. This scenario
leads naturally to the problem of coupled
resistively shunted junctions which has been studied in the
literature in different contexts \cite{rsj}.
Our arguments, therefore, follow the ones used in
the problem of granular superconductors
\cite{grain}, and one can show that the network of stripes undergoes a
Kosterlitz-Thouless (KT) phase transition towards a superconducting
state at a critical temperature $T_c$. We also show how
doping with Zn changes this picture in order to drive the
system towards a superconducting-insulating phase transition
at an universal value of the resistance given by
$R_Q = h/(4 e^2) \approx 6.45 k \Omega$.

Indeed, in a recent paper Fukuzumi {\it et al.} \cite{zync} studied the
temperature behavior
of the resistivity of single crystals of
YBa$_2$(Cu$_{1-z}$Zn$_z$)$_3$O$_{7-y}$
and La$_{2-x}$Sr$_x$Cu$_{1-z}$Zn$_z$O$_4$.
It was shown that these materials undergo a universal superconducting-insulator
transition as a function of the Zn doping, and indeed 
the superconducting critical temperature in the underdoped samples
seems to vanish very close to the universal value $R_Q$.
This behavior is reminiscent of the
behavior in granular superconductors where the resistivity, instead
of decreasing close to the superconducting transition, actually
increases if the sample resistance is greater than $R_Q$ \cite{grain}.

At zero temperature we assume the network to be in a superconducting
state which is characterized by a gap $|\Delta|$ and a superconducting
phase $\Phi$.
We also assume that the connection between the stripes and the lakes is
a perfect interface.
The electronic system on the stripes is
described in terms of a Luttinger liquid \cite{euro}.
The Luttinger liquid is described in terms
of right, R, and left, L, moving fermions with spin
$\sigma = \uparrow,\downarrow$
which are created (destroyed)
by operators $\psi^{\dag}_{R,L,\sigma}(x)$ ($\psi_{R,L,\sigma}(x)$). These
fermions can be bosonized via the transformation \cite{boso}
$
\psi_{R,L,\sigma}(x) = \sqrt{k_F} e^{\pm i \sqrt{\pi}
\phi_{R,L,\sigma}(x)}
$.
The bosonic modes
$\phi$ can be described in terms of amplitude, $\phi_{\alpha}$,
and phase, $\theta_{\alpha}$, modes as
$
\phi_{R,L,\alpha}(y) = \phi_{\alpha}(y) \mp \theta_{\alpha}(y) .
$
In turn these bosonic fields can be written in terms
of charge and spin bosonic modes,
$
\phi_{\rho,s} = (\phi_{\uparrow} \pm \phi_{\downarrow})/\sqrt{2}$
and
$
\theta_{\rho,s} = (\theta_{\uparrow} \pm \theta_{\downarrow})/\sqrt{2}
$,
and it is easy to show that the Euclidean Lagrangean density
of the system can be written as (with the units $\hbar=k_B=1$),
\begin{eqnarray}
{\cal L}_S = \sum_{i=\rho,s} \left\{\frac{g_i}{2 v_i}
\left[\left(\partial_{\tau} \phi_{i}\right)^2 + v_{i}^2
\left(\partial_x \phi_{i}\right)^2\right]\right\}.
\label{ac}
\end{eqnarray}
$g_s$ and $g_{\rho} $are the Luttinger parameters
for spin and charge respectively, and $v_s$ and $v_{\rho}$ are
their velocities.

It is believed that the stripes fluctuate in the superconducting
phase. Thus, the electrons on the stripe will
undergo strong backscattering (by corners, for instance) which can
lead to localization.
Since we assume that the stripes are
metallic, we have to rely on a strong attractive interaction between
the electrons in order to get delocalization. This attraction
could be provided, for instance, by
the surrounding antiferromagnet \cite{emery,new} and the phase
coherence by the mechanism described in this paper. 
Renormalization group studies of disordered Luttinger
liquids show that if $g_{\rho}$ is smaller than a critical value
$g_c$ (for singlet pairing $g_c^s = 1/3$), the Luttinger liquid delocalizes.
In this case it was shown in ref.\cite{gs} that the temperature
dependence of the resistance is given by,
\begin{equation}
R(T) \approx T^{1+\gamma}
\label{res}
\end{equation}
where  $\gamma =1/g_{\rho}-1/g_c$.
This result is also consistent with the presence of a Cooper-pair gap in
the normal phase of these materials. Our picture is the one
where in the normal state of these materials the electrons
on the stripe are paired but there is no superconducting
phase coherence. As it was explained by Emery and Kivelson
\cite{steve}, this is possible in one dimension because
pairing and phase coherence have completely different origins.
This could be an explanation of the so-called ``spin gap" \cite{spingap}
seen in some cuprates.
Since the conduction occurs within the stripes, the only effect of Zn
is to add a residual zero temperature resistance, $R_0$ (besides
the effects it can have on a d-wave order parameter \cite{pines}).
This leads to a total resistance,
$
R_s(T) = R_0 + A  T^{1+\gamma}
$
where $A$ is a non-universal coefficient which {\it does not}
depend on Zn doping but on the stripe fluctuations. Indeed in
the experiments the linear part of resistance is insensitive
to Zn doping \cite{zync}.

At zero
temperature the system is in a superconducting state
and a Josephson coupling
develops between the Luttinger stripes via the lakes.
At finite temperatures, above the bulk superconducting critical
temperature, $T_c$, the
electrons propagate in the system via the network of Luttinger liquids
but there is no phase coherence.
At low temperatures the scattering is dominated by the Luttinger liquid
scattering since the quasi-two dimensional scattering is much smaller.
Experimentally it has been established for quite some time that the
resistivity behaves linearly with temperature.
Since the cuprates are very close to a superconductor
to insulator transition one has from the above discussion that
$\gamma \ll 1$ which is consistent with this linear behavior.
Deviations from the linear power have
also been observed \cite{nakano} and can be absorbed
into $\gamma$. At low temperatures as the Cooper pairs propagate 
they are going to feel differences in the superconducting gap
in different regions.
This is known to lead to Andreev reflections \cite{and}.
Therefore one has to add to the Luttinger liquid Lagrangean
(\ref{ac}) another term which is related to the presence of
the gap. This problem is very similar to the problem of
superconducting metals coupled by a Luttinger liquid that
has been studied in the context of mesoscopic physics \cite{urb}.
The Lagrangean associated with pairing is
\begin{eqnarray}
{\cal L}_{pair} &=& |\Delta| \left\{\cos\left[\Phi + \sqrt{2 \pi}
(\theta_{\rho}
- \phi_{s})\right] \right.
\nonumber
\\
&+& \left. \cos\left[\Phi+ \sqrt{2 \pi} (\theta_{\rho}
+ \phi_{s})\right]\right\}.
\nonumber
\end{eqnarray}
If $|\Delta|$ is large, the main effect of the cosine term is
to pin the value of the fields to the minima of the potential, i.e.,
\begin{eqnarray}
\langle \theta_{\rho} \rangle &=& \left(\pi n - \Phi\right)/(\sqrt{2 \pi})
\label{bound}
\end{eqnarray}
where $n$ is an integer.
This implies that in the regions where the gap is large the bosonic
fields are subjected to twisted boundary conditions \cite{urb}.

We can calculate the current density that flows in the stripes which
is given by 
$
j(x) = -2 v_{\rho}\partial_x \theta_{\rho}/(\sqrt{2 \pi}g_{\rho})$.
According to (\ref{bound}) it is given by
\begin{equation}
j(x) = v_{\rho} \partial_x \Phi/(\pi g_{\rho}).
\label{jp}
\end{equation}
This is the value of the supercurrent density at zero temperature. As
expected it depends only on the value of the phase. Moreover,
our argument is based on a large value of the gap which is
only valid at zero temperature. At finite temperatures the phase
fluctuates. In
this case we have to add
another contribution to the problem which is the normal dissipative
current.
Thus, if the distance between two stripes is $\ell$ (see Fig.(\ref{imp}))
and $\Phi$ is the phase difference between them
we have from (\ref{jp}) that the Josephson current is
\begin{equation}
I_{J}(\Phi) = e v_{\rho} \Phi/(\pi g_{\rho} \ell)
\label{jozero}
\end{equation}
where $e$ is the electronic charge.
Notice that the energy scale in the problem is given by
\begin{eqnarray}
T_L = v_{\rho}/\ell \, ,
\label{tl}
\end{eqnarray}
and the Josephson energy in this case is
\begin{equation}
E_J = I_J(2 \pi)/(2 e) = T_L/g_{\rho}.
\label{energy}
\end{equation}

The total current
from one stripe to another has three
different contributions: the Josephson current, a normal
dissipative current, and a current through
a capacitor with capacitance $C$ which corresponds to the charge
accumulation at the lakes.
First we consider the problem of two stripes connected by one
lake. Henceforth we introduce the phase as a quantum
mechanical operator which is conjugated to the number of Cooper pairs
($[\Phi,n]=i$). The Josephson effect is described by
the potential $V(\Phi) = E_J U(\Phi)$
where $U(\Phi) = \Phi^2/2$ (module $2\pi$)
is the potential associated with (\ref{jozero}).
The simplest way to mimic the dissipative part of the Luttinger-stripe
is to introduce a set of decoupled harmonic oscillators
following Caldeira-Leggett. The Lagrangean is \cite{thesis}
\begin{eqnarray}
{\cal L} &=& \frac{M}{2} \left(\frac{d \Phi}{d t}\right)^2 + E_J U(\Phi)
\nonumber
\\
&-& \frac{1}{2 e} \frac{d \Phi}{d t} \sum_i \lambda_i x_i +
\sum_i \left(\frac{p_i^2}{2 m_i} + \frac{1}{2} m_i \omega_i^2 x_i^2\right),
\label{la}
\end{eqnarray}
where $M = C/(2 e)^2$. This Lagrangean describes the motion
of a fictitious particle with mass $M$ moving in a periodic potential
$U$ coupled to a heat bath.
As was shown before \cite{thesis}, the properties of this system
depend only on the spectral function
$J(\omega)= \frac{\pi}{2} \sum_i (\lambda_i^2 \omega_i/m_i)
\delta(\omega-\omega_i)$.
By requiring that the voltage between the lakes is given by the resistance
$R_S$ times the normal current we find that this spectral function
is uniquely given by
$J(\omega,T) = \omega/R_s(T)$.
It is well established that the model described by (\ref{la}) has a
zero temperature phase transition as a function of the parameter
$\alpha(T) = R_Q/R_s(T)$. It has been shown \cite{thesis} that
for $\alpha(0) <1$ the junction has a finite resistance and for
$\alpha(0)>1$ the junction is in a superconducting state with zero
resistance.
This model has a duality symmetry \cite{tb} which allows
the calculation of the resistance as a function of temperature
at low temperatures ($T<<T_s=1/\sqrt{A C}$).
One has
\begin{equation}
\frac{R(T)}{R_Q} \approx\frac{\Gamma[\alpha(T)]
\pi^{2 \alpha(T)+1/2}}{2 \Gamma[\alpha(T)+1/2]}
\left(\frac{E_J}{\gamma(T)}\right)^2 \left(\frac{T}{\gamma(T)}\right)^{
2(\alpha(T)-1)}
\nonumber
\end{equation}
where $\gamma(T) = 1/(R_s(T) C)$ and $\Gamma[z]$ is a Gamma function.

In the absence of Zn we have $R_0 \approx 0$ \cite{zync}. 
This implies that
the resistance behaves like
$R(T) \approx 2 R_Q e^{- 2 T_A/T \ln(T_s/T)}$ where $T_A = R_Q/A$ 
and is rapidly suppressed at
low temperatures showing the growth of superconducting fluctuations
in the system. However, superconductivity is obtained only at zero
temperature because we have only two connected stripes.
In the Zn-doped case one can substitute for $\alpha(T)$
its zero temperature value at low temperatures,
$\alpha_0 =R_Q/R_0$, and one finds
that
$
R(T) \sim T^{2 (\alpha_0-1)}.
$
Therefore, for $\alpha_0>1$ ($R_0<R_Q$) the resistance goes
to zero and superconductivity is obtained. For $\alpha_0<1$ ($R_0>R_Q$)
the resistance becomes very large at low temperatures indicating
a transition to an insulating state. Moreover, for $R_0 \approx R_Q$ one 
finds a logarithmic behavior. These results are consistent with the available 
data on Zn doped cuprates \cite{zync}.

In order to explain the finite temperature phase transition one has to
rely on the geometric structure of the array of lakes and Luttinger
stripe rivers. In order to do that we generalize the Lagrangean
(\ref{la}) to ${\cal L} = \sum_{a,b} {\cal L}_{ab}(\Phi_{ab})$,
where, ${\cal L}_{ab}$ is given in (\ref{la}) and $a,b$ label each two stripes
linked by one lake. The calculation of the partition function for
the problem is analogous to the one for two stripes. However we have to
introduce the disordered distribution of lakes with different sizes and
stripes with different lengths. One can show that this is
a problem of a stack of X-Y models coupled in the imaginary time
direction with random couplings \cite{thesis}.  We can then coarse
grain the imaginary time direction and reduce the system to one
effective {\it classical} X-Y model with renormalized coupling
constants.  When this is done, one finds a partition function for a
classical model given by $Z = \int D\phi \, \, \, \exp
\left\{-\sum_{ab} \kappa_{ab} (\phi_a-\phi_b)^2/2
 \right\}$ where at low temperatures in the pure case
($\alpha(T) \to \infty$ as $T \to 0$) we have $ \kappa(T) \approx E_J/T $.
This leads naturally to a KT transition at some critical value of the coupling,
$\kappa_c$ which depends on the detailed structure of the lattice
\cite{kt}. If $\kappa<\kappa_c$ then the system is disordered with a
exponentially decaying correlation function.  For $\kappa>\kappa_c$ the
correlation function decays as a power law, and we have quasi-long range
order. Since $\kappa(T)$ diverges at low temperatures for $\alpha_0 >1$, the
system has a transition to quasi-long range order at some critical
temperature $T_c$, which is given by (using (\ref{tl}) and (\ref{energy})),
\begin{equation}
T_c = T_L/(g_{\rho} \kappa_c)
= v_{\rho}/(g_{\rho} \kappa_c \ell).
\label{super}
\end{equation}
This critical temperature signals the transition
to a superconducting state with vanishing resistance.
Observe that $\kappa_c$ depends on the impurities. Highly
disconnected lattices (which can be caused by Zn doping) have
large $\kappa_c$ that decreases substantially $T_c$. 

One of the most interesting results of  Yamada {\it et al.} \cite{yamada}
is the observed relation between $T_c$ and the incommensurability
$\epsilon$ which is found to be
$
T_c(K) \approx 181 \epsilon(\AA^{-1})
$
for a wide range of doping.
In our picture one can relate $T_c$ to
the inter-stripe distance using $\epsilon = \pi/\ell$. Thus,
{\it from the experiments} one concludes that
\begin{eqnarray}
T_c \approx 569/\ell.
\label{tc}
\end{eqnarray}
By comparing (\ref{super}) and (\ref{tc}),
we find perfect agreement between experiment and theory.
Moreover, from (\ref{tc}), assuming
$g_{\rho}=1/3$ (which gives $R_s(T) \propto T$) and $\kappa_c \approx
0.35$ (for the square lattice)
\cite{kt} one finds (restoring $\hbar$ and $k_B$) $\hbar v_{\rho}
\approx 6. 10^{-2} eV \AA$. We point out that indications of
KT transition have been observed in
YBa$_2$Cu$_3$O$_{7-\delta}$ and LaBaCuO \cite{ybco}. 
Another non-trivial but straightforward
prediction of this theory is the behavior of superconducting coherence
length $\xi_s$ with doping. Since the maximum value of the phase gradient
in the theory is $\pi/(2 \ell)$ a simple Ginzburg-Landau argument 
gives $\xi_s \sim \ell$ \cite{tinkham}. Thus, in the clean limit one expects the
upper critical field $H_{c2}$ to behave like 
$H_{c2} \propto 1/\ell^2$. Comparing this prediction with neutron scattering
data in La$_{2-x}$Sr$_x$CuO$_{4}$ 
\cite{yamada} one finds $H_{c2} \propto T_c^2$ over a large
doping region ($0.06 \leq x \leq 0.2$). 
In particular for small doping levels ($0.06 \leq x \leq 0.12$)
one finds $H_{c2} \propto x^2$. These predictions can be tested in
new measurements of $H_{c2}$ at low temperatures in high magnetic fields
\cite{boebinger}.

In conclusion, we propose a new scenario for the
problem of phase coherence and superconductivity in striped
cuprates which is based on the assumption of a network of
Luttinger-stripes
and lakes of holes in the CuO$_2$ planes. In our model the stripes
are pinned by impurities and
a Josephson current is transferred from stripe to stripe via the
network. The same network also carries a normal
current which dissipates and is responsible for the
superconducting-insulator transition seen in these materials.
In our case the superconducting transition is a two-dimensional 
KT transition and our model is consistent with many existing experimental
data, especially the recent neutron scattering experiments by
Yamada {\it et al.} \cite{yamada}.
It also explains the recently discovered universal behavior
of the superconducting-insulator transition in the presence
of Zn impurities \cite{zync}. Moreover, we predict the behavior
of $H_{c2}$ as a function of $T_c$ and doping, $x$.

I thank A.Balatsky, W. Beyermann, G. Castilla, D.Hone, S. Kivelson,
D. MacLaughlin, E.Miranda, C.Nayak, L.Radzihovsky, C.Smith and
J.Tranquada for useful discussions and comments.

{\bf Note}: After this paper was submitted I became aware of the
experimental paper in ref.\cite{ultimo} where the increase of
disconnectivity of the network was observed in $\mu$SR under
Zn doping and a related work on Luttinger liquid networks \cite{paco}.

\end{document}